\documentclass[final,5p,times,twocolumn]{elsarticle}
\usepackage{graphicx}
\usepackage{amsmath,amssymb,amsfonts}
\usepackage[english]{babel}
\usepackage[normalem]{ulem}
\usepackage[dvipsnames]{xcolor}
\usepackage{siunitx} 
\usepackage{hyperref}

\biboptions{sort&compress}

\hypersetup{
    colorlinks=true,
    linkcolor=blue,
    citecolor=blue,
    urlcolor=blue
}

\journal{Mechanics of Materials}

\begin{document}


\title{High-throughput characterization of snap-through stability boundaries \\
of bistable beams in a programmable rotating platform}

\author[EPFL]{E.~Gutierrez-Prieto}
\author[EPFL]{G.~Yakir}
\author[EPFL]{P.M.~Reis\corref{cor1}}

\address[EPFL]{Flexible Structures Laboratory,\\ 
Institute of Mechanical Engineering,\\
École Polytechnique fédérale de Lausanne (EPFL),\\
1015 Lausanne, Switzerland}

\cortext[cor1]{Corresponding author}
\ead{pedro.reis@epfl.ch}

\begin{abstract}
We introduce a high-throughput platform that enables simultaneous, parallel testing of six bistable beams via programmable motion of a rotating disk. By prescribing harmonic angular dynamics, the platform explores the phase space of angular velocity and acceleration $(\Omega,\,\dot{\Omega})$, producing continuously varying centrifugal and Euler force fields that act as tunable body forces in our specimens. Image processing extracts beam kinematics with sub-pixel accuracy, enabling precise identification of snap-through events. By testing six beams in parallel, the platform allows systematic variation of beam thickness, pre-compression, tilt angle, and clamp orientations across 65 distinct configurations, generating 23,400 individual experiments. We construct stability boundaries and quantitatively parameterize them as parabolic functions, characterized by a vertical offset and a curvature parameter. Tilt angle provides the most robust mechanism for tuning the curvature parameter, while beam thickness and pre-compression modulate vertical offset. Modal decomposition analysis reveals that antisymmetric clamp configurations can trigger mode switching, in which competing geometric and inertial effects drive transitions through different deformation pathways.
Our work establishes a scalable experimental framework for high-throughput characterization of dynamic nonlinear instabilities in mechanics. The complete experimental dataset is made publicly available to support data-driven design and machine learning models for nonlinear mechanics with applications to bistability-based metamaterials, mechanical memory, and electronics-free sensing systems.
\end{abstract}

\maketitle


\section{Introduction}

Experimental mechanics has traditionally relied on sequential testing protocols that examine one specimen at a time. This approach, while yielding precise measurements, constrains the exploration of parameter spaces to relatively small datasets. High-throughput methodologies have transformed materials discovery and characterization~\cite{szymanski2023autonomous,schneider2018automating}. Recent efforts have started to demonstrate automated testing strategies in mechanics. For example, robotized test rigs have been employed for advanced material model discovery on metals ~\cite{tancogne2024using,roth2025quantifying}. Integration of FEM, additive manufacturing, and tensile tests has permitted optimization of auxetic structures~\cite{gongora2021using,meier2024obtaining} and rapid development of soft robots~\cite{hardman2025automated,chen2020design}. However, these efforts often require specialized equipment for each configuration, preventing experimental parallelization and limiting comprehensive exploration of systems with many degrees of freedom.

Multistable mechanical structures would benefit from high-throughput characterization. These structures exploit multiple local energy minima to achieve distinct stable configurations without continuous energy input, enabling applications in energy absorption~\cite{cao2021bistable,shan2015multistable}, shape reconfiguration~\cite{fu2018morphable,haghpanah2016multistable,yang2019multi}, haptic displays~\cite{abbasi2024leveraging}, and mechanical information storage~\cite{chen2021reprogrammable,gutierrez2025dynamic,kwakernaak2023counting,yang2018multistable}. While static analyses of canonical bistable elements such as the von~Mises truss and pre-arched beams guide understanding~\cite{MisesZAMM1923,BelliniJNLM1972,qiu2004curved,zhao2008post,medina2014experimental,pandey2014dynamics,gomez2017critical,gomez2019dynamics,abbasi2023snap,chen2023snap,gutierrez2024harnessing}, systematic experimental mapping of their dynamic behavior across parameter spaces remains elusive. This limitation hinders the rational design of arrays with many multistable elements, such as systems with many mechanical bits~\cite{harne2017harnessing,kwakernaak2023counting,gutierrez2025dynamic}. Conventional experimental protocols render exhaustive parameter characterization impractical, while computational approaches capable of capturing snap-through dynamics without quasi-static assumptions require computationally expensive resources~\cite{wilson1972nonlinear,jr2004iterative,tong2020microtubule}.

A promising avenue for high-throughput characterization of bistable structures emerges from the use of unsteady rotations. In previous work~\cite{gutierrez2024harnessing,gutierrez2025dynamic}, we demonstrated how centrifugal forces, $\mathbf{f}_{\Omega}$, and Euler forces, $\mathbf{f}_{E}$, arising from controlled angular motions can induce snap-through in rotating bistable beams. These non-inertial forces are dynamically coupled via time derivatives. 
The loading state at time $t$ is characterized by the pair of values $\{\Omega(t),\,\dot{\Omega}(t)\}$, where $\Omega(t)$ and $\dot{\Omega}(t)=\text{d}\Omega(t)/\text{d}t$ denote the instantaneous angular velocity and acceleration.

The stability boundaries of such beams are naturally described in the $(\Omega,\,\dot{\Omega})$ phase space, where snap-through transitions occur when the loading trajectory crosses critical thresholds that form approximately parabolic curves~\cite{gutierrez2025dynamic}.
By prescribing $\Omega(t)$ and $\dot{\Omega}(t)$, we can systematically explore this phase space and map the thresholds for the up-switching and down-switching events, $\{\Omega^{\uparrow}$, $\dot{\Omega}^{\uparrow}\}$ and $\{\Omega^{\downarrow}$, $\dot{\Omega}^{\downarrow}\}$, respectively. 
By tailoring the boundary conditions and geometry of individual beams, selective actuation within an array becomes possible even if all beams experience the same global rotational loading~\cite{gutierrez2025dynamic}. Extensive experimental mapping of how these stability boundaries vary across the parameter space of geometric and boundary conditions remains unaddressed.

Here, we introduce a high-throughput experimental platform that exploits programmable non-inertial loading to characterize bistable beam stability boundaries. The platform tests six specimens simultaneously, each subjected to hundreds of loading trajectories that systematically explore the $(\Omega,\,\dot{\Omega})$ phase space through controlled disk rotation. By varying beam thickness, pre-compression, tilt angle, and clamp orientations across 65 configurations, we generate 23,400 experiments.
This dataset represents an unprecedented volume of experimental data for dynamic instability phenomena in mechanics, enabling statistical characterization and providing a foundation database for future data-driven modeling approaches. We quantitatively parameterize the resulting stability boundaries as parabolic functions, revealing how geometric and boundary conditions systematically tune the snap-through thresholds. 

\begin{figure}[b!]
    \includegraphics[width=\linewidth]{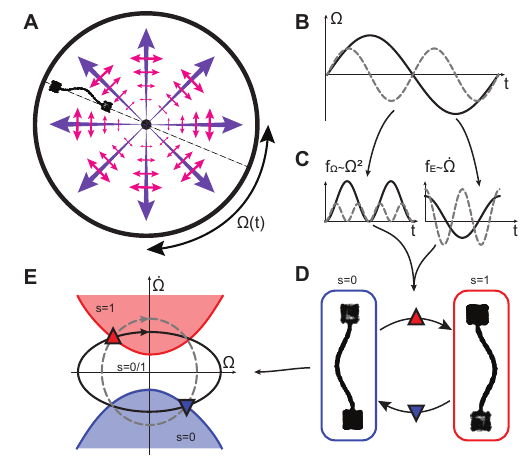}
    \caption{\textbf{High-throughput characterization of stability boundaries via programmable unsteady rotation}. \textbf{A.} Schematic of the rotating platform showing centrifugal $\mathbf{f}_{\Omega}$ (purple) and Euler $\mathbf{f}_{E}$ (pink) force distributions acting on a bistable beam. \textbf{B.} Imposed angular velocity profiles $\Omega(t)$ of varying amplitude produce \textbf{C.} time-varying forces scaling as $\mathbf{f}_{\Omega}\sim \Omega^2$ (centrifugal) and $\mathbf{f}_{E}\sim \dot{\Omega}$ (Euler). \textbf{D.} Snap-through transitions between stable states ($s=0$ and $s=1$) occur when loading trajectories cross stability thresholds. \textbf{E.} Varying loading profiles trace elliptical orbits in the $(\Omega,\,\dot{\Omega})$ phase space, serving to map the stability boundaries (shaded regions).}
    \label{fig:concept}
\end{figure}

Figure~\ref{fig:concept} presents the general approach of our rotating platform.
A bistable beam positioned on a rotating disk (Figure~\ref{fig:concept}A) experiences time-varying non-inertial forces when subjected to prescribed angular velocity profiles $\Omega(t)$ of varying amplitude (Figure~\ref{fig:concept}B--C). These forces trigger cyclic snap-through events between stable states ($s=0$ and $s=1$) when loading trajectories cross stability boundary thresholds (Figure~\ref{fig:concept}D). By performing multiple cyclic loadings of varying amplitude and frequency, which trace elliptical orbits in the $(\Omega,\,\dot{\Omega})$ phase space, we map the switching thresholds (Figure~\ref{fig:concept}E). The contact-free non-inertial loading enables simultaneous testing of multiple specimens under precisely controlled, time-varying loads, replacing sequential testing protocols with parallelized experiments that probe the entire $(\Omega,\,\dot{\Omega})$ parameter space. We compile a high-volume database of experimental data and make it publicly available to support future machine-learning applications and data-driven design in nonlinear mechanics~\cite{db_placeholder}.

This manuscript is organized as follows: Section~\ref{sec:methods} describes the high-throughput rotational platform; Section~\ref{sec:results} presents a parametric study examining how stability boundaries vary across configurations; and Section~\ref{sec:modes} presents a modal decomposition analysis to reveal deformation pathways during snap-through. Finally, in Section~\ref{sec:discusion_conclusion}, we discuss the insights gained and the implications of our study, summarize the main findings, and provide an outlook for future work.

\section{High-throughput experimental platform and methods}
\label{sec:methods}

We develop a high-throughput characterization rotating platform that generates programmable non-inertial force fields via controlled angular motions. This section describes the platform architecture (Section~\ref{sec:platform_design}), non-inertial loading approach (Section~\ref{sec:force_fields}), image processing (Section~\ref{sec:imageprocessing}), and statistical protocol for constructing stability boundaries (Section~\ref{sec:protocol}).

\subsection{Apparatus design, specimen fabrication, and parameters}
\label{sec:platform_design}

\begin{figure}[h!]
    \centering
    \includegraphics[width=0.8\linewidth]{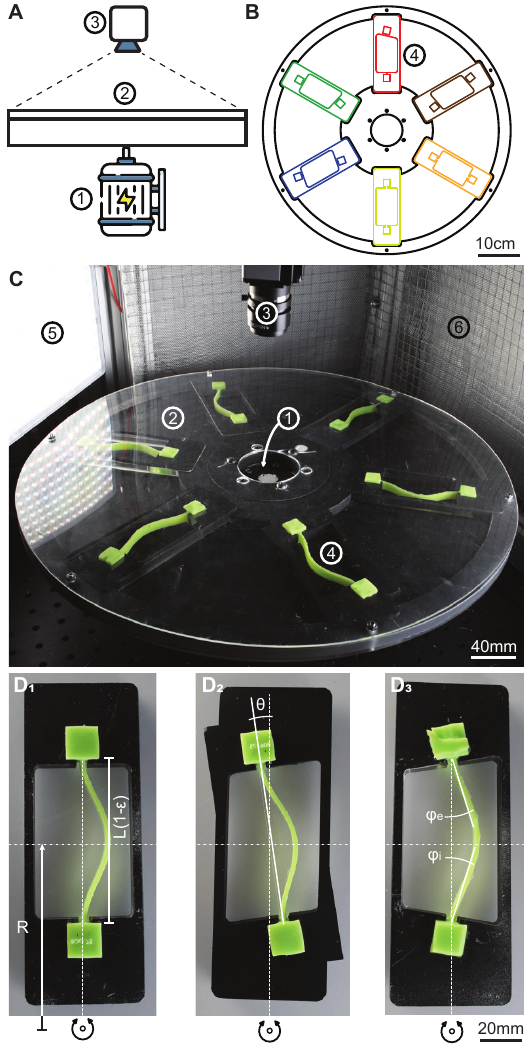}
    \caption{\textbf{Experimental apparatus}. \textbf{A.} Side- and \textbf{B.} top-view schematics.
    \textbf{C.} Photograph of the apparatus. A high-torque motor (1) rotates a disk (2) with grooves for up to six sample holders (3). A camera (4) records the disk from above, illuminated by LED panels (5), all inside a safety enclosure (6). \textbf{D.} Modular sample holders enable independent control of (\textbf{D$_1$.}) pre-compression $\varepsilon$ and radial position $R$, (\textbf{D$_2$.}) tilt angle $\theta$, and (\textbf{D$_3$.}) clamp orientations $\varphi_i$, $\varphi_e$.}
    \label{fig:setup}
\end{figure}

Figure~\ref{fig:setup} presents our rotating platform with the schematics in Figure~\ref{fig:setup}A--B illustrating the core components: a direct-drive motor (1) rotates a disk (2), onto which we mount the specimens, which are imaged by an overhead camera (3) in the lab frame. The disk features six evenly spaced radial grooves that accommodate custom-designed sample holders (4), enabling simultaneous testing of up to six distinct beams and different boundary conditions in a single experimental run.

\paragraph{Apparatus}
The assembled system shown in Figure~\ref{fig:setup}C comprises a high-torque direct-drive motor (1) (ETEL RTMBi140-050-RBS), providing angular control with $\pm 15$ arcsec resolution, that rotates a hollow acrylic disk (2) with 300\,mm radius and 15\,mm thickness, within which we place the specimens mounted in the sample holders (4). The system imposes angular profiles within maximal values $|\Omega_{\text{max}}| \lesssim 100\,\text{rad\,s}^{-1}$ and $|\dot{\Omega}_{\text{max}}| \lesssim 1000\,\text{rad\,s}^{-2}$. A global-shutter digital camera (3) (IDS U3-3040SE-M-GL with Vision-Lens 3.5\,mm EFL, f/1.4) mounted in the lab frame records at up to 700\,fps with 320$\times$320 pixel resolution. Two high-intensity LED panels (5) provide illumination, enabling reduced shutter speeds (100\,$\mu$s) to prevent motion blur. A stainless steel mesh enclosure (6) surrounds the setup for safety.

\paragraph{Fabrication of beams}
The beams are fabricated with a silicone-based elastomer (VPS32, Zhermack Elite Double 32) with Young's modulus $E = 1.22 \pm 0.05\,\mathrm{MPa}$, Poisson's ratio $\nu\approx 0.5$, and density $\rho = 1.17 \pm 0.01\cdot 10^3\, \mathrm{kg}\, \mathrm{m}^{-3}$~\cite{GrandgeorgeJMPS2022,LeroyPRF2022}. Injection molding with laser-cut acrylic molds produces uniform rectangular beams with thickness $h \in [1.5,\,3.0]\,\mathrm{mm}$, length $L = 80\,\mathrm{mm}$, and width $b = 8\,\mathrm{mm}$. The beam ends are cast into larger rectangular blocks ($15\times15\times8\,\text{mm}^{3}$) for clamping. 

\paragraph{Parameters and boundary conditions}
The modular sample holders (Figure~\ref{fig:setup}D) enable systematic variation of pre-compression $\varepsilon$, tilt angle $\theta$, and clamp orientations $\varphi_i$ (radially internal) and $\varphi_e$ (external). These parameters can be used to set symmetric, antisymmetric, or single-clamp configurations. The system's modularity enables rapid resetting between experimental runs, significantly accelerating parametric exploration. Table~\ref{tab:bcs} summarizes all the parameters explored in this study.

\begin{table*}[h!]
\centering
\caption{\textbf{Sets of explored geometric parameters and boundary conditions.}}
\label{tab:bcs}
\begin{tabular}{llll}
\hline
\textbf{Description} & \textbf{Parameter} & \textbf{Units} &\textbf{Values} \\
\hline
Thickness & $h$ & mm & $1.60, 1.75, 2.10, 2.35, 2.50, 2.70$\\
Pre-compression & $\varepsilon$ & - & $0.015,0.020,0.025,0.030,0.035,0.040,0.045,0.050,0.060$\\
Tilt angle & $\theta$ & deg & $0.0,\pm2.5,\pm5.0,\pm10.0,\pm15.0,\pm20.0$\\
Symmetric clamp angles & $\varphi$ ($\varphi_i=\varphi_e$) & deg & $0.0,\pm2.5,\pm5.0,\pm7.5,\pm10.0,\pm12.5,\pm15.0$ \\
Antisymmetric clamp angles & $\varphi$ ($\varphi_i=-\varphi_e$) & deg & $0.0,\pm2.5,\pm5.0,\pm7.5,\pm10.0,\pm12.5,\pm15.0$\\
Internal clamp angle & $\varphi$ ($\varphi_e=0$) & deg & $0.0,\pm5.0,\pm10.0,\pm15.0$\\
External clamp angle & $\varphi$ ($\varphi_i=0$) & deg & $\pm5.0,\pm10.0,\pm15.0$\\[3pt]
\hline
\end{tabular}
\end{table*}

\subsection{Programmable loading and phase-space exploration}
\label{sec:force_fields}

In the frame of reference of the rotating disk, a bistable beam of linear density $\rho$ and located at radius $R$ (measured at the midpoint between clamps) experiences three non-inertial forces~\cite{landau1976Mechanics,feynman1965feynman,gutierrez2024harnessing}: the centrifugal force, $\mathbf{f}_{\Omega} = \rho R\Omega^2 \mathbf{e}_r ds$; the Euler force, $\mathbf{f}_{E} = -\rho R\dot{\Omega} \mathbf{e}_t ds$; and the Coriolis force, $\mathbf{f}_{C} = -2\rho(\Omega\mathbf{e}_z \times \mathbf{v})ds$, where $\mathbf{v}$ is the relative velocity with respect to the rotating frame, $\{\mathbf{e}_r,\,\mathbf{e}_t,\,\mathbf{e}_z\}$ are the orthogonal base vectors in the radial, tangential and vertical directions, and $\text{d}s$ is a differential line segment. For beams rigidly attached to the disk, $\mathbf{v}$ is negligible except during rapid snap-through, and thus $\mathbf{f}_{C}$ can be disregarded~\cite{gutierrez2024harnessing}. The beam dynamics are governed by the pair of orthogonal forces,  $\mathbf{f}_{\Omega}$ and $\mathbf{f}_{E}$.

These centrifugal and Euler forces play distinct roles: $\mathbf{f}_{\Omega}$ is conservative and modifies the beam's effective potential-energy landscape, whereas $\mathbf{f}_{E}$ is non-conservative and can exert work during transitions. An elastic beam pre-compressed to an end-to-end shortening $\varepsilon$ buckles into an arch whose internal energy features two distinct minima, each corresponding to a stable state. Varying $\Omega$ alters the centrifugal potential to affect the energy levels and can introduce asymmetries between the stable states. During accelerations, $\mathbf{f}_{E}$ provides the work required to overcome the potential energy barrier triggering snap-through.

Coupled through the time derivative, the joint action of $\mathbf{f}_{\Omega}$ and $\mathbf{f}_{E}$ establishes the $(\Omega,\,\dot{\Omega})$ phase space as the natural domain for characterizing system stability~\cite{gutierrez2025dynamic}. 
In this phase space, a stability boundary separates the regions where only one state is stable from those where both states can coexist, as sketched in Figure~\ref{fig:concept}E. Physically, this boundary represents the loading state at which the energy barrier between the two stable configurations is overcome. To map these stability boundaries, we prescribe harmonic angular velocity profiles of the form
\begin{equation}
\label{eq:harmonic}
    \Omega(t) = a \sin(2\pi f t),
\end{equation}
where $a$ and $f$ are the drive amplitude and frequency. The maximum magnitudes of the forces are
\begin{equation}
\begin{split}
    |\mathbf{f}_{\Omega,\text{max}}| &= \rho R a^2,\\
    |\mathbf{f}_{E,\text{max}}| &= 2\pi \rho R f a,
\end{split}
    \label{eq:f_max}
\end{equation}
implying that varying amplitude $a$ affects both $\mathbf{f}_{\Omega}$ and $\mathbf{f}_{E}$, whereas varying frequency $f$ modulates only $\mathbf{f}_{E}$. As illustrated in Figure~\ref{fig:concept}E, harmonic drives trace elliptical orbits in the $(\Omega,\,\dot{\Omega})$ phase space. 
To explore the phase-space extensively, as detailed in Table~\ref{tab:loads}, we vary $f$ across four values ($f = \{0.75,\,1.50,\,2.25,\,3.00\}\,\mathrm{Hz}$) and $a$ across fifteen values ranging from $a=5.0$ to approximately $80.0\,\mathrm{rad\,s}^{-1}$, with the exact range adjusted per frequency to remain within motor capabilities. This method systematically probes the two-dimensional loading space and identifies the specific combinations that induce snap-through transitions~\cite{gutierrez2025dynamic}.

\begin{table*}[h!]
\centering
\caption{\textbf{Loading parameters.} 
}
\label{tab:loads}
\begin{tabular}{r|l|c|c}
\hline
Frequency [Hz] & Amplitude [rad s$^{-1}$] & $s_0$ & Repetitions\\
\hline
0.75 & 5.0, 10.4, 15.7, 21.1, 26.4, 31.8, 37.1, 42.5, 47.9, 53.2, 58.6, 63.9, 69.3, 74.6, 80.0 & 0, 1 & 3\\[3pt]
1.50 & 5.0, 10.4, 15.7, 21.1, 26.4, 31.8, 37.1, 42.5, 47.9, 53.2, 58.6, 63.9, 69.3, 74.6, 80.0 & 0, 1 & 3\\[3pt]
2.25 & 5.0, 9.7, 14.4, 19.1, 23.8, 28.5, 33.2, 37.9, 42.6, 47.3, 52.0, 56.6, 61.3, 66.0, 70.7 & 0, 1 & 3\\[3pt]
3.00 & 5.0, 8.4, 11.9, 15.3, 18.7, 22.2, 25.6, 29.0, 32.5, 35.9, 39.3, 42.8, 46.2, 49.6, 53.1 & 0, 1 & 3\\[3pt]
\hline
\end{tabular}
\end{table*}

\subsection{Image processing and definition of snap-through events}
\label{sec:imageprocessing}

\begin{figure}[h!]
   \centering
   \includegraphics[width=\linewidth]{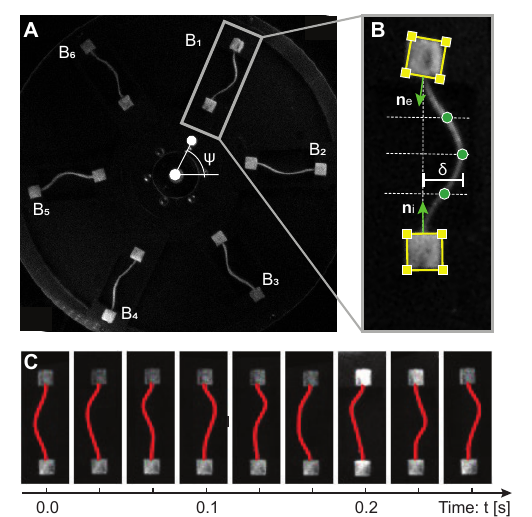}
   \caption{
   \textbf{Image processing pipeline.} 
    \textbf{A.} Representative raw experimental image. White circles serve as fiducial markers, and their centroids define the instantaneous rotation angle $\psi$.
    \textbf{B.} Schematic of the detection and spline-fitting workflow. A neural network locates the clamp corners (yellow squares), from which the beam's clamp-to-clamp axis (dashed vertical line) and the normal vectors (green arrows) to the internal and external clamps, $\mathbf{n}_i$ and $\mathbf{n}_e$, are established. Three search lines, perpendicular to the clamp-to-clamp axis, detect intensity maxima corresponding to the spline interpolation nodes (green circles). 
    \textbf{C.} Time-sequence montage of fitted splines (red solid lines) overlaid on raw beam images exhibits excellent agreement.
   }\label{fig:processing}
\end{figure}

\begin{figure*}[h!]
   \centering
   \includegraphics[width=\linewidth]{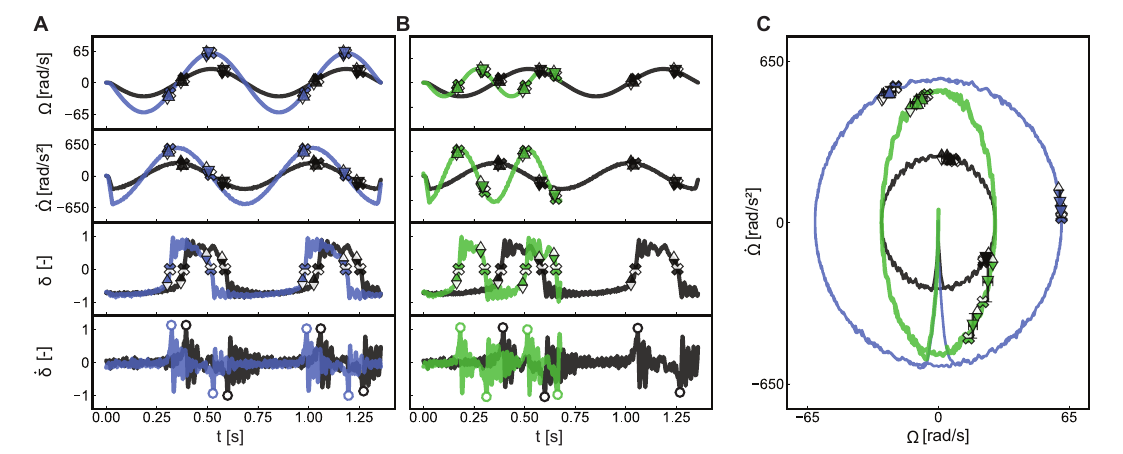}
   \caption{
   \textbf{Definition of snap-through events.} 
    \textbf{A--B.} From top to bottom, measurements of angular velocity $\Omega(t)$, angular acceleration $\dot{\Omega}(t)$, midspan displacement $\delta(t)$, and midspan velocity $\dot{\delta}(t)$ of a representative bistable beam ($h=1.85\,\mathrm{mm}$, $\varepsilon=0.035$, $\theta=5^\circ$) for two harmonic loadings of varying (\textbf{A}) amplitude ($f=1.5\,\mathrm{Hz}$ with black lines: $a=32.5\,\mathrm{rad\,s}^{-1}$; blue lines: $a=65.0\,\mathrm{rad\,s}^{-1}$), and (\textbf{B}) frequency ($a=32.5\,\mathrm{rad\,s}^{-1}$ with black lines: $f=1.5\,\mathrm{Hz}$; green lines: $f=3.0\,\mathrm{Hz}$). Snap-through events (triangles) are defined as the mean value between the point where $\delta=0$ (crosses) and when the beam midspan reaches velocity $|\dot{\delta}| = 0.15\cdot|\dot{\delta}_{\mathrm{Max}}|$ (diamonds). The snapping direction is indicated by triangle orientation ($s=0{\rightarrow} s=1$: up-triangle; $s=1{\rightarrow} s=0$: down-triangle).
    \textbf{C.} Representation of the three harmonic drives from panels \textbf{A} (black and blue lines) and \textbf{B} (black and green lines) in the $(\Omega,\,\dot{\Omega})$ phase space. Snapping points are overlaid for each loading orbit.
   }\label{fig:snap_definition}
\end{figure*}

The automated data-processing pipeline extracts beam kinematics from recorded images, as illustrated in Figure~\ref{fig:processing}. The algorithm first determines the disk's instantaneous angle from fiducial markers, as shown in the raw image in Figure~\ref{fig:processing}A. We validate this measurement against the motor-encoder output to confirm synchronization, but we avoid image rotation to prevent quality loss during interpolation.

For each beam, a pre-trained fully convolutional neural network (U-Net for semantic segmentation and localization) locates the clamp corners (yellow squares in Figure~\ref{fig:processing}B). This network generates probability heatmaps, and the final coordinates are determined by identifying local maxima. The network was trained using the \texttt{PyTorch} library~\cite{paszke2017automatic} on manually annotated images spanning various lighting conditions, beam orientations, and clamp geometries, with data augmentation including rotation, brightness variation, and noise distortion. The identified clamp corners define the beam's clamp-to-clamp axis (dashed vertical line), along which we define the longitudinal coordinate $x$ (ranging from $x=0$ at the internal clamp to $x=1$ at the external one). The network also establishes the orientation vectors $\mathbf{n}_{i}$ and $\mathbf{n}_{e}$ of the internal and external clamps, respectively (green arrows). The beam centerline is identified at sub-pixel resolution (approximately $\pm0.2\,\mathrm{mm}$) by fitting Gaussian functions to brightness profiles along search lines perpendicular to the axis (dashed horizontal lines). We reconstruct the full beam shape by fitting a Hermite exponential spline to the detected centerline points, $\mathbf{n}_{i}$ and $\mathbf{n}_{e}$. Fig~\ref{fig:processing}C shows a time series of splines (red) superimposed on raw experimental images, demonstrating excellent agreement. The $C^2$-continuous Hermite splines permit calculation of local curvature, which may be of interest for future applications of the dataset.

We compute the beam’s deflection, $w(x,t)$, defined as the transverse component of the centerline position with respect to the clamp-to-clamp axis. From the spline series, we extract the centerline as a position vector $\mathbf{r}(s,\,t)$, where $s$ represents arc length (from $s=0$ at the internal clamp to $s=L$ at the external one). We also extract the beam's midspan displacement defined as $\delta(t)=w(L/2,t)$; see Fig~\ref{fig:processing}B. This displacement metric serves as the primary state indicator: the sign of $\delta$ distinguishes between the two stable configurations, with $\delta>0$ corresponding to clockwise buckling (state $s=1$, see Figure~\ref{fig:concept}D) and $\delta<0$ to counterclockwise buckling (state $s=0$).

Defining the precise moment of snap-through requires an empirical choice. While a transition is often described as occurring at the instant when the beam's midpoint crosses the neutral axis ($\delta=0$), this occurs when the snap-through is already well underway and irreversible. The onset of instability is a more physically relevant marker. Therefore, we define each snap-through event as a bounded interval. As illustrated by time series of $\delta(t)$ and $\dot{\delta}(t)$ in Figure~\ref{fig:snap_definition}A--B, we identify two bounds: the zero-crossing point (crosses) and the snap onset, defined when the midspan velocity $|\dot{\delta}|$ first reaches 15\% of its peak value during that transition (hollow circles: peak velocity; diamonds: empirical snap onset). The snap-through event is recorded as the mean point of this range (triangles), with error bars spanning the onset-to-crossing interval. This way we identify the critical snapping points for the up-switching ($\Omega^{\uparrow},\,\dot{\Omega}^{\uparrow}$) and down-switching ($\Omega^{\downarrow},\,\dot{\Omega}^{\downarrow}$) events. 

When observing snap-through events in the time domain (Figure~\ref{fig:snap_definition}A--B), no simple loading threshold is apparent. Modulating amplitude $a$ (Figure~\ref{fig:snap_definition}A) or frequency $f$ (Figure~\ref{fig:snap_definition}B) causes snap-through events to occur at different instantaneous values of $\Omega$ and $\dot{\Omega}$. From this temporal analysis, the transition points appear disordered and do not reveal a clear switching threshold.

An underlying structure emerges when the data is plotted in the $(\Omega,\,\dot{\Omega})$ phase space, as introduced in our previous work~\cite{gutierrez2025dynamic}. Figure~\ref{fig:snap_definition}C presents this phase-space representation, where prescribed harmonic drives trace elliptical orbits. Modifying the amplitude produces concentric ellipses (black and blue orbits), whereas adjusting the frequency affects only vertical elongation (black and green orbits). When snap-through events are mapped onto this phase space, they fall within specific regions rather than appearing randomly scattered. This representation demonstrates that beam stability is determined not by a single loading parameter, but by the specific combination of $\Omega$ and $\dot{\Omega}$ that governs the coupled loading pair of $\mathbf{f}_{\Omega}$ and $\mathbf{f}_{E}$. However, the snap-through events exhibit substantial spread within these regions, arising from experimental uncertainties in geometry, boundary conditions, and loading execution. This variability calls for a statistical approach to robustly characterize the stability boundaries.

\subsection{Statistical construction of stability boundaries}
\label{sec:protocol}

To characterize stability boundaries, we probe the $(\Omega,\,\dot{\Omega})$ phase space by imposing families of harmonic angular velocity profiles described by Eq.~\ref{eq:harmonic}, whose size and eccentricity are controlled by $a$ and $f$. For each $(f,\,a)$ pair, experiments are conducted starting from both initial states ($s_0=0$ and $s_0=1$) with three repetitions per loading condition, running a total of 360 experiments per specimen and generating between 200 and 600 individual snap-through events per set of boundary conditions (see Table~\ref{tab:loads}).

\begin{figure}[h!]
\centering
\includegraphics[width=\linewidth]{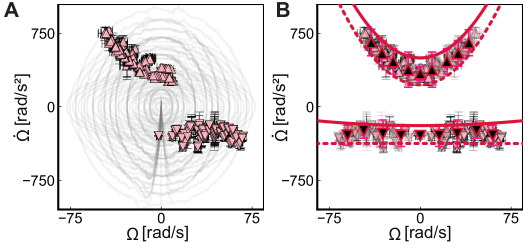}
    \caption{
    \textbf{Statistical construction of stability boundaries in the $(\Omega,\,\dot{\Omega})$ phase space.}
    \textbf{A.} A set of loading orbits (light gray lines) generates a dense, structured cloud of snap-through events (pink triangles) for a representative specimen ($h=1.85\,\mathrm{mm}$, $\varepsilon=0.035$, $\theta=5^\circ$).
    \textbf{B.} Snap-through data are mirrored about $\Omega=0$ to construct the full boundary. Parabolic functions (solid and dashed black lines) are fitted to the binned data (black circles with error bars) to define a 95\% confidence region for the stability boundary.
}\label{fig:curves}
\end{figure}

As illustrated in Figure~\ref{fig:curves}A, the collected snap-through events form a dense, structured cloud of points for each transition direction. This cloud arises from combined experimental uncertainties, including fabrication tolerances in beam geometry and uncertainties in imposed boundary conditions and executed loading profiles. Despite variability, the data points delineate a clear boundary representing the instability threshold: when a loading trajectory crosses this boundary, the beam's current configuration becomes unstable, triggering an irreversible transition to an alternative state.

Following Eq.~\ref{eq:f_max}, $\mathbf{f}_{\Omega}$ is symmetric about $\Omega=0$, and thus stability boundaries must also be symmetric. We implement this physical symmetry in Figure~\ref{fig:curves}B by mirroring the collected data points about $\Omega=0$ to construct the complete boundary for each transition. We quantitatively characterize these boundaries using parabolic functions
\begin{equation}
    \dot{\Omega}(\Omega) = C_0 + C_2\Omega^2,
\label{eq:parabola}
\end{equation}
where the vertical offset $C_0$ corresponds to the stability threshold at zero angular velocity, and $C_2$ represents the curvature parameter of the parabola. We use a parabolic approximation because it is the simplest functional form that captures the symmetry and nonlinearity of the empirical data, without implying a fundamental physical derivation of the shape.

To fit the quadratic polynomial to the scattered data, we employ a statistical binning approach. For each transition direction, we divide the mirrored data into nine equally spaced bins of width $\Delta\Omega = 15\,\mathrm{rad\,s}^{-1}$ along the $\Omega$-axis and compute the mean value $\langle\dot{\Omega}\rangle$ and standard deviation $S_{\dot{\Omega}}$ of the snap-through events within each bin, as illustrated in Figure~\ref{fig:curves}B. We then fit two parabolas, one to the upper bounds $\langle\dot{\Omega}\rangle + 2S_{\dot{\Omega}}$ (solid lines) and one to the lower bounds $\langle\dot{\Omega}\rangle - 2S_{\dot{\Omega}}$ (dashed lines) of the binned data for each snap direction, defining a 95\% confidence interval for the stability boundary.

\section{Parametric study of stability boundaries}
\label{sec:results}

Having established the experimental platform and methodology, we now present results from a comprehensive parametric study. We systematically characterize how beam thickness $h$, pre-compression $\varepsilon$, tilt angle $\theta$, and clamp orientations $\varphi_i$ and $\varphi_e$ affect the snap-through stability boundaries. By tracking the evolution of the parabolic coefficients $C_0$ and $C_2$ introduced in Eq.~\ref{eq:parabola} across parameter space, we quantitatively map their influence and provide design guidelines for tuning dynamic thresholds. Table~\ref{tab:bcs} presented above summarizes all 65 configurations examined, spanning over 23,400 independent experiments.

\begin{figure}[h!]
    \centering
    \includegraphics[width=\linewidth]{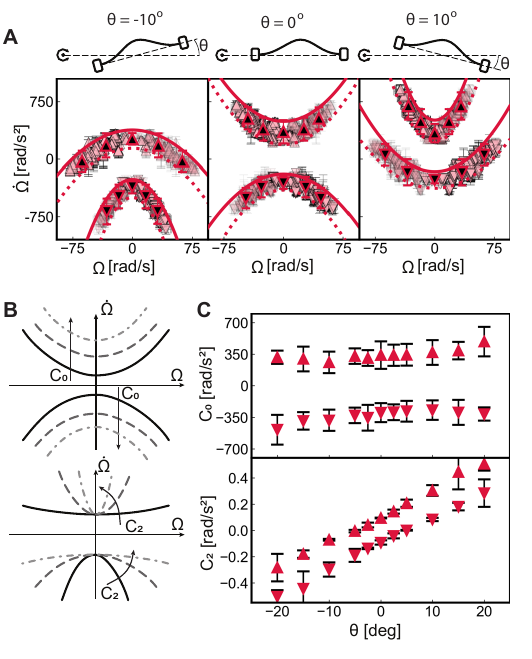}
    \caption{
    \textbf{Effect of tilt angle on the stability-boundary coefficients.} 
    \textbf{A.} Measured stability boundaries for beams tilted at $\theta=-10^\circ$, $0^\circ$, and $10^\circ$. Sketches show beam orientation. Upward triangles denote $0{\rightarrow} 1$ transitions, downward triangles denote $1{\rightarrow} 0$ transitions. \textbf{B.} Schematics illustrate how increasing $C_0$ shifts boundaries vertically (top) while increasing $C_2$ modulates curvature (bottom). \textbf{C.} Evolution of the parabolic coefficients $C_0$ (top) and $C_2$ (bottom) versus tilt angle $\theta$. Markers show mean values with error bars 
    representing 95\% confidence intervals from the parabolic fits.}\label{fig:tilt}
\end{figure}

\paragraph{Effect of tilt angle} We first investigate the effect of tilting the beam's longitudinal axis by an angle $\theta$ with respect to the radial direction. Figure~\ref{fig:tilt}A shows stability boundaries for $\theta=\{-10^{\circ},0^{\circ},10^{\circ}\}$. For the aligned case ($\theta=0^\circ$, central panel), boundaries for up-switching ($0{\rightarrow}1$) and down-switching ($1{\rightarrow}0$) are highly symmetric. Introducing the tilt angle breaks this symmetry (left and right panels), altering the parabolic curvature in opposing ways depending on the sign of $\theta$. In Figure~\ref{fig:tilt}B, we schematize the effect of $C_1$ and $C_2$ on the fitted parabolas. In Figure~\ref{fig:tilt}C, we quantify how the coefficients are affected by $\theta$; $C_0$ remains largely unaffected, while $C_2$ exhibits strong linear dependence on $\theta$. This behavior arises from force decomposition: non-zero $\theta$ causes the radial centrifugal force $\mathbf{f}_{\Omega}$ to have a transverse component proportional to $|\mathbf{f}_{\Omega}|\sin\theta$. This transverse component, which depends on $\Omega^2$, acts as a bias favoring one buckled state while opposing the other. The tilt angle thus modulates the curvature parameter $C_2$ without significantly affecting the vertical offset $C_0$.

\begin{figure}[h!]
\centering
\includegraphics[width=\linewidth]{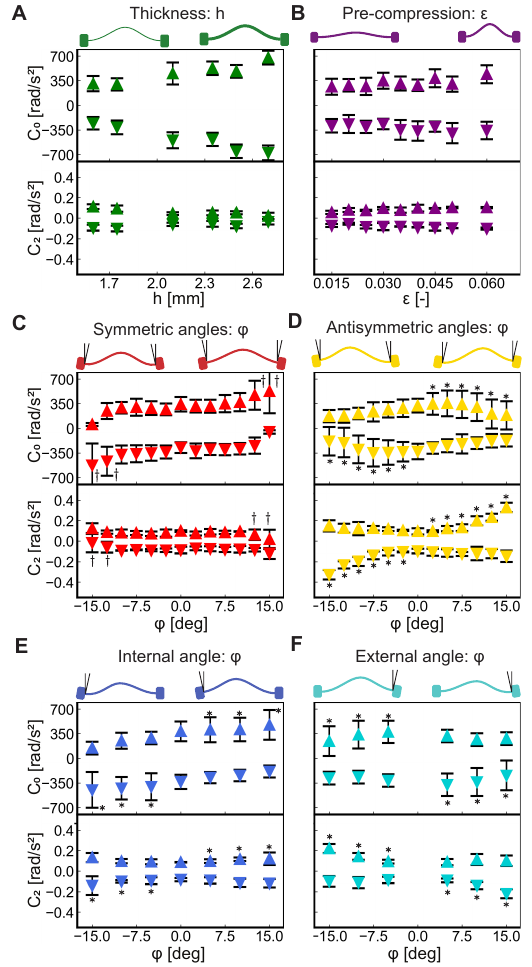}
    \caption{
    \textbf{Parametric effects on $C_0$ and $C_2$.} Evolution of parabolic coefficients $C_0$ (top row) and $C_2$ (bottom row) as functions of (\textbf{A}) beam thickness $h$, (\textbf{B}) pre-compression $\varepsilon$, (\textbf{C}) symmetric clamp angles $\varphi=\varphi_i=\varphi_e$, (\textbf{D}) antisymmetric angles $\varphi=\varphi_i=-\varphi_e$, (\textbf{E}) internal clamp angle $\varphi_i$ with $\varphi_e=0$, and (\textbf{F}) external angle $\varphi_e$ with $\varphi_i=0$. Upward triangles denote $0{\rightarrow}1$ transitions, downward triangles denote $1{\rightarrow}0$ transitions. Error bars represent 95\% confidence intervals. Daggers ($\dagger$) mark configurations where one stable state is nearly eliminated; asterisks ($\ast$) indicate non-parabolic boundaries (see text).
}\label{fig:results_summary}
\end{figure}

\paragraph{Effect of beam thickness}
As observed in Figure~\ref{fig:results_summary}A, increasing $h$ (thus, bending stiffness, $EI\sim h^3$) leads to a monotonic increase in the magnitude of $C_0$ without significantly affecting $C_2$. The energy barrier separating the two stable states scales with $EI$, requiring greater work from the $\mathbf{f}_{E}$ to trigger snap-through. Across the tested range ($h\in[1.6,\,2.7]\,\mathrm{mm}$), $C_0$ increases approximately linearly with $h$ for both switching directions, with similar slope magnitudes. The $C_2$ parameter remains largely unchanged because thickness does not break the system's radial symmetry; both stable states experience identical stiffness increases. Beam thickness thus provides a direct design parameter for scaling snap-through forces without altering the phase-space topology or introducing directional asymmetries.

\paragraph{Effect of pre-compression}
A similar stiffening effect, although more modest, is observed when increasing $\varepsilon$, as shown in Figure~\ref{fig:results_summary}B. Pre-compression raises the elastic energy stored in the buckled arch, increasing the barrier to snap-through. Across the tested range ($\varepsilon\in[0.015,\,0.060]$), $C_0$ increases monotonically with $\varepsilon$ for both switching directions, though less pronounced than with $h$. Like thickness, pre-compression preserves radial symmetry, leaving $C_2$ relatively unchanged, offering a complementary tuning mechanism: it modulates thresholds through geometric prestress rather than material stiffness, enabling independent control when combined with thickness.

\paragraph{Effect of clamp angles}
(i) Symmetrically angling the clamps ($\varphi_i = \varphi_e = \varphi$) has minimal effect on stability boundaries for moderate angles (Figure~\ref{fig:results_summary}C). Only at the most extreme angles tested (marked with $\dagger$) does a significant shift occur, but one stable state is nearly eliminated at these limits, leading to substantial uncertainties in the fitted coefficients. (ii) Antisymmetric clamp angles ($\varphi_i = -\varphi_e = \varphi$) produce far more pronounced effects (Figure~\ref{fig:results_summary}D). The coefficients exhibit non-monotonic trends, and at large angles, the boundaries deviate from a parabolic form, with distinct kinks. The parabolic model becomes inadequate for these cases (marked with $\ast$), discussed further in Section~\ref{sec:modes}. When only one clamp angle is varied, complementary behaviors emerge. (iii) Modifying the internal angle $\varphi_i$ (with $\varphi_e=0$) primarily affects $C_0$ with minimal impact on $C_2$ (Figure~\ref{fig:results_summary}E). Conversely, varying only the external angle $\varphi_e$ (with $\varphi_i=0$) significantly changes $C_2$ while leaving $C_0$ largely unaffected (Figure~\ref{fig:results_summary}F). This behavior enables independent tuning of threshold and curvature parameters. Some extreme configurations exhibit non-parabolic boundaries (marked with $\ast$).

The results from this parametric study reveal complementary tuning mechanisms: thickness and pre-compression uniformly shift thresholds ($C_0$) while preserving symmetry, whereas tilt and clamp angles introduce asymmetries that modulate curvature ($C_2$). The independent control provided by internal and external clamp angles enables the precise design of stability landscapes. The non-parabolic boundaries observed at extreme angles indicate transitions in deformation mode, which we examine through modal analysis in Section~\ref{sec:modes} to elucidate the physical mechanisms underlying complex stability topologies.

\section{Mode switching in snap-through transitions}
\label{sec:modes}

Thus far, our analysis has characterized snap-through dynamics using $\delta(t)$, the beam's midpoint deflection, but this single metric cannot explain the non-parabolic features observed in Section~\ref{sec:results}. To examine the underlying mechanism, we decompose the beam's centerline deflection $w(x,\,t)$ into a Fourier series
\begin{equation}
    w(x,t) = \sum_{i=1}^{N} A_i(t) \phi_i(x),
\label{eq:fourier}
\end{equation}
where $A_i(t)$ are the time-dependent modal amplitudes and $\phi_i(x)$ are the spatial basis functions. We focus on the first two modes, which capture the essential physics of the transition. The first symmetric mode, $A_1$, corresponds to the primary buckling shape
\begin{equation}
w_1(x,t) = A_1(t)\cos\left(\frac{\pi x}{L}\right),
\label{eq:mode1}
\end{equation}
and the first antisymmetric mode, $A_2$, captures the unstable S-shaped deformation that mediates snap-through
\begin{equation}
w_2(x,t) = A_2(t)\sin\left(\frac{\pi x}{L}\right).
\label{eq:mode2}
\end{equation}

\begin{figure}[t!]
\centering
\includegraphics[width=\linewidth]{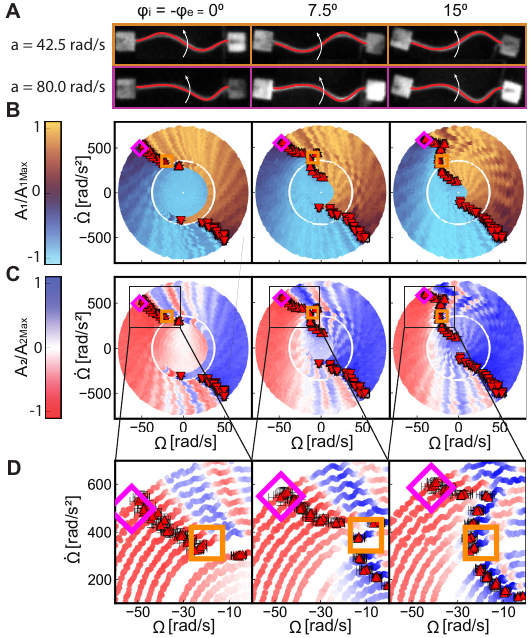}
\caption{
    \textbf{Modal decomposition of snap-through transitions.}
    \textbf{A.} Experimental snapshots during snap-through for two loading amplitudes ($a=\{42.5,\,80.0\}\,\mathrm{rad/s}$, rows) and three antisymmetric clamp angles ($\varphi=\{0^\circ,\,7.5^\circ,\,15^\circ\}$, columns). Line overlays indicate the centerline obtained via image processing.
    \textbf{B--C.} Normalized modal amplitudes in the $(\Omega,\dot{\Omega})$ phase space for (\textbf{B}) first symmetric mode $A_1$ and (\textbf{C}) first antisymmetric mode $A_2$, shown for $\varphi=\{0^\circ,\,7.5^\circ,\,15^\circ\}$ (columns). Color scales span $[-1,\,1]$ normalized to maximum amplitude. Triangles mark snap-through events (up: $0{\rightarrow} 1$, down: $1{\rightarrow} 0$). The kink in the up-switching boundary ($\varphi=\{7.5^\circ,\,15^\circ\}$) corresponds to regions where $A_2$ changes sign. \textbf{D.} Magnified views of the kink regions, with colored boxes (pink and orange) matching the snapshots in panel \textbf{A}, for $a=\{42.5,\,80.0\}\,\mathrm{rad/s}$, respectively.
}\label{fig:modes}
\end{figure}

Figure~\ref{fig:modes}A shows snapshots during snap-through for two loading amplitudes (rows) and three antisymmetric clamp angles (columns), with overlaid centerlines from image processing. These images confirm that the antisymmetric mode $A_2$ mediating the transition varies with loading conditions and geometric configuration. For a beam without angled clamps ($\varphi=0^\circ$, left column), snap-through proceeds via the same S-shaped deformation pathway regardless of loading amplitude; the unstable intermediate shape maintains the same orientation (pink and orange markers in panels B--D correspond to the low and high amplitude transitions, respectively). In contrast, for beams with antisymmetric clamps (center column: $\varphi=7.5^\circ$; right column: $\varphi=15^\circ$), changing the loading amplitude causes the unstable S-shaped mode to flip orientation. This mode switching manifests as distinct intermediate deformation shapes during the transition.

We quantify this mode switching by plotting $A_1$ and $A_2$ as heatmaps in the $(\Omega,\,\dot{\Omega})$ phase space (Figure~\ref{fig:modes}B--D). The symmetric mode $A_1$ transitions sharply from positive to negative values at the stability boundaries, marked by overlaid triangles in Figure~\ref{fig:modes}B. For antisymmetric clamps, the up-switching boundary exhibits a clear kink, whereas the down-switching boundary remains parabolic. Analysis of $A_2$ in Figure~\ref{fig:modes}C--D reveals the cause. Without angled clamps ($\varphi=0^\circ$), transitions always occur with negative $A_2$ amplitude. With antisymmetric clamps ($\varphi\neq0$), however, snap-through can proceed via either positive or negative $A_2$, depending on the location in phase space. The kink marks the conditions at which the snap-through pathway switches.

This mode switching arises from competition between the geometric and the non-inertial loading effects. For positive antisymmetric angles ($\varphi > 0$), the angled clamps favor snap-through via the antisymmetric mode with negative $A_2$ amplitude. Simultaneously, the centrifugal force acting on a beam in state $s=0$  pushes the point of maximum deflection radially outward, favoring positive $A_2$ amplitude. These two effects are in direct competition: at low $\Omega$, the geometric bias dominates, and the beam snaps via negative $A_2$; at high $\Omega$, centrifugal loading overcomes this bias, driving snap-through via positive $A_2$ instead. The kink in the stability boundary marks the transition between these regimes.

This competition vanishes during down-switching ($1 {\rightarrow} 0$). For a beam in state $s=1$, the centrifugal force again pushes the point of maximum deflection radially outward, now corresponding to negative $A_2$ amplitude. Both the clamp geometry and the non-inertial loading favor the same deformation mode, leading to constructive effects and a boundary without kinks. 

This analysis reveals the physical origin of the non-parabolic boundaries reported in Section~\ref{sec:results} for some special cases. Mode switching occurs when geometric and non-inertial effects compete, producing stability boundaries that the simple two-coefficient parabolic model cannot capture.

\section{Discussion and conclusion}
\label{sec:discusion_conclusion}

Our systematic investigation demonstrates that the stability boundaries of bistable beams in a rotating frame, under non-inertial loading, can be accurately described, albeit empirically, by parabolic functions in the $(\Omega,\,\dot{\Omega})$ phase space across a broad range of geometric configurations. The vertical offset $C_0$ and curvature parameter $C_2$ provide quantifiable metrics for characterizing these boundaries. Parameter ranking reveals that tilt angle $\theta$ provides the most robust mechanism for asymmetric tuning, linearly modulating $C_2$. Beam thickness $h$ and pre-compression $\varepsilon$ effectively modulate energy barriers through $C_0$, with $h$ exhibiting a stronger influence than $\varepsilon$. Clamp angles provide additional control: internal angles primarily affect $C_0$, whereas external angles predominantly tune $C_2$, thereby providing complementary tuning of both parameters.

Modal decomposition reveals that antisymmetric clamp configurations induce mode-switching in which the snap-through pathway varies with the loading trajectory and boundary conditions. In these configurations, geometric bias from angled clamps competes with centrifugal loading, causing the antisymmetric mode $A_2$ to change sign with angular velocity. At low $\Omega$, geometric effects dominate; at high $\Omega$, centrifugal forcing prevails. This competition explains the non-parabolic kinks observed in certain stability boundaries and demonstrates that a single metric, such as the midpoint displacement $\delta(t)$, cannot fully capture the complexity of these dynamic transitions. The parabolic description remains valid when geometric and non-inertial effects act constructively, as occurs for symmetric configurations.

This framework provides design guidelines for dynamically addressable bistable arrays. In practice, beam thickness $h$ sets the overall energy scale, pre-compression $\varepsilon$ offers fine-tuning of the vertical offset, and tilt angle $\theta$ adjusts the parabolic curvature of the boundaries. Clamp angles offer additional degrees of freedom for creating complex boundary topologies. This parametric control enables the design of intersecting stability boundaries required for multi-bit mechanical memory systems, where individual elements must be selectively actuated under global loading conditions~\cite{gutierrez2025dynamic}. Beyond mechanical memory applications, the identification of mode switching phenomena opens avenues for designing mechanical devices sensitive to deformation pathways during snap-through.

The high-throughput experimental methodology demonstrated here can be extended to other classes of nonlinear mechanical systems to accelerate parameter space exploration and model validation. Our comprehensive dataset, comprising over 23,400 individual experiments with detailed kinematic information, is made publicly available~\cite{db_placeholder} to support future efforts on data-driven modeling and machine learning approaches for controlling nonlinear dynamical phenomena. Possible extensions of the present study include more complex loading profiles, material nonlinearities (\textit{e.g.}, viscoplasticity), and functional structures involving bistable shells or metamaterial integration. This approach promises to accelerate the design and optimization of adaptive, multifunctional mechanical systems.

\bigskip

\noindent \textbf{Acknowledgments.} The authors acknowledge the EPFL Center for Imaging, especially Dr. Florian Aymanns, for assistance/help with developing the image processing pipeline. The authors also acknowledge C.M. Meulblok and M.~van Hecke for insightful discussions on the content of this manuscript.

\noindent \textbf{Competing interests.} The authors declare no competing interests.

\noindent \textbf{Author contributions.} All authors contributed to the idea creation, writing and editing of the manuscript. E.G.P. and G.Y. developed the experimental setup and produced the experimental data. E.G.P. developed the control and data processing algorithms, produced the figures, and wrote the first draft of the manuscript.

\noindent \textbf{Author Information.} The authors declare no competing financial interests. Correspondence and requests for materials should be addressed to P.M.R. (pedro.reis@epfl.ch).

\noindent \textbf{Data Availability} The database supporting the findings of this study is available in the Zenodo public repository \texttt{[placeholderlink]}~\cite{db_placeholder}.

\noindent\textbf{Declaration of Generative AI Usage}. During the preparation of this work, the authors used Claude.ai, Gemini.google.com, and Grammarly.com to improve the text's quality and ensure consistency in terminology. After using this tool/service, the authors reviewed and edited the content as needed and take full responsibility for the content of the published article.

\bibliography{references}

@article{cao2021bistable,
  title = {Bistable structures for advanced functional systems},
  author = {Cao, Yunteng and Derakhshani, Masoud and Fang, Yuhui and Huang, Guoliang and Cao, Changyong},
  journal = {Adv. Funct. Mater.},
  volume = {31},
  number = {45},
  pages = {2106231},
  year = {2021}
}

@article{gomez2019dynamics,
  title = {Dynamics of viscoelastic snap-through},
  author = {Gomez, Michael and Moulton, Derek E and Vella, Dominic},
  journal = {J. Mech. Phys. Solids},
  volume = {124},
  pages = {781--813},
  year = {2019}
}

@article{chen2021reprogrammable,
  title = {A reprogrammable mechanical metamaterial with stable memory},
  author = {Chen, Tian and Pauly, Mark and Reis, Pedro M},
  journal = {Nature},
  volume = {589},
  number = {7842},
  pages = {386--390},
  year = {2021}
}

@article{qiu2004curved,
  title = {A curved-beam bistable mechanism},
  author = {Qiu, Jin and Lang, Jeffrey H and Slocum, Alexander H},
  journal = {J. Microelectromech. Syst.},
  volume = {13},
  number = {2},
  pages = {137--146},
  year = {2004}
}

@article{gomez2017critical,
  title = {Critical slowing down in purely elastic ‘snap-through’ instabilities},
  author = {Gomez, Michael and Moulton, Derek E and Vella, Dominic},
  journal = {Nat. Phys.},
  volume = {13},
  number = {2},
  pages = {142--145},
  year = {2017}
}

@article{zhao2008post,
  title = {Post-buckling and snap-through behavior of inclined slender beams},
  author = {Zhao, Jian and Jia, Jianyuan and He, Xiaoping and Wang, Hongxi},
  journal = {J. Appl. Mech.},
  year = {2008}
}

@article{abbasi2023snap,
  title = {Snap buckling of bistable beams under combined mechanical and magnetic loading},
  author = {Abbasi, Arefeh and Sano, Tomohiko G and Yan, Dong and Reis, Pedro M},
  journal = {Philos. Trans. R. Soc. A},
  volume = {381},
  number = {2244},
  pages = {20220029},
  year = {2023}
}

@article{chen2023snap,
  title = {Snap-through dynamics of a buckled flexible filament with different edge conditions},
  author = {Chen, Zepeng and Mao, Qian and Liu, Yingzheng and Sung, Hyung Jin},
  journal = {Phys. Fluids},
  volume = {35},
  number = {10},
  year = {2023}
}

@article{kwakernaak2023counting,
  title = {Counting and sequential information processing in mechanical metamaterials},
  author = {Kwakernaak, Lennard J and van Hecke, Martin},
  journal = {Phys. Rev. Lett.},
  volume = {130},
  number = {26},
  pages = {268204},
  year = {2023}
}

@book{harne2017harnessing,
  title = {Harnessing bistable structural dynamics: for vibration control, energy harvesting and sensing},
  author = {Harne, Ryan L and Wang, Kon-Well},
  year = {2017},
  publisher = {John Wiley \& Sons}
}

@article{szymanski2023autonomous,
  title = {An autonomous laboratory for the accelerated synthesis of novel materials},
  author = {Szymanski, Nathan J and Rendy, Bernardus and Fei, Yuxing and Kumar, Rishi E and He, Tanjin and others},
  journal = {Nature},
  volume = {624},
  number = {7990},
  pages = {86--91},
  year = {2023}
}

@article{schneider2018automating,
  title = {Automating drug discovery},
  author = {Schneider, Gisbert},
  journal = {Nat. Rev. Drug Discov.},
  volume = {17},
  number = {2},
  pages = {97--113},
  year = {2018}
}

@article{wilson1972nonlinear,
  title = {Nonlinear dynamic analysis of complex structures},
  author = {Wilson, EL and Farhoomand, I and Bathe, KJ},
  journal = {Earthq. Eng. Struct. Dyn.},
  volume = {1},
  number = {3},
  pages = {241--252},
  year = {1972}
}

@article{jr2004iterative,
  title = {Iterative coupling of {BEM} and {FEM} for nonlinear dynamic analyses},
  author = {Jr, D Soares and Estorff, O von and Mansur, WJ},
  journal = {Comput. Mech.},
  volume = {34},
  number = {1},
  pages = {67--73},
  year = {2004}
}

@article{gutierrez2024harnessing,
  title = {Harnessing centrifugal and {Euler} forces for tunable buckling of a rotating elastica},
  author = {Gutierrez-Prieto, Eduardo and Gomez, Michael and Reis, Pedro M},
  journal = {Extreme Mech. Lett.},
  volume = {72},
  pages = {102246},
  year = {2024}
}

@article{gutierrez2025dynamic,
  title = {Dynamic driving enables independent control of material bits for targeted memory},
  author = {Gutierrez-Prieto, E and Meulblok, CM and van Hecke, M and Reis, PM},
  journal = {arXiv preprint arXiv:2508.16257},
  year = {2025}
}

@article{MisesZAMM1923,
  title = {\"Uber die {S}tabilitätsprobleme der {E}lastizit\"atstheorie},
  author = {Mises, R. V.},
  journal = {Z. Angew. Math. Mech.},
  volume = {3},
  number = {6},
  pages = {406--422},
  year = {1923}
}

@article{BelliniJNLM1972,
  title = {The concept of snap-buckling illustrated by a simple model},
  author = {Paul X. Bellini},
  journal = {Int. J. Non-Linear Mech.},
  volume = {7},
  number = {6},
  pages = {643--650},
  year = {1972}
}

@article{GrandgeorgeJMPS2022,
  title = {An elastic rod in frictional contact with a rigid cylinder},
  author = {Grandgeorge, Paul and Sano, Tomohiko G. and Reis, Pedro M.},
  journal = {J. Mech. Phys. Solids},
  volume = {164},
  pages = {104885},
  year = {2022}
}

@article{LeroyPRF2022,
  title = {Tapered foils favor traveling-wave kinematics to enhance the performance of flapping propulsion},
  author = {Leroy-Calatayud, Pierre and Pezzulla, Matteo and Keiser, Armelle and Mulleners, Karen and Reis, Pedro M.},
  journal = {Phys. Rev. Fluids},
  volume = {7},
  number = {7},
  pages = {074403},
  year = {2022}
}

@article{shan2015multistable,
  title = {Multistable architected materials for trapping elastic strain energy},
  author = {Shan, Sicong and Kang, Sung H and Raney, Jordan R and Wang, Pai and Fang, Lichen and Candido, Francisco and Lewis, Jennifer A and Bertoldi, Katia},
  journal = {Adv. Mater.},
  volume = {27},
  number = {29},
  pages = {4296--4301},
  year = {2015}
}

@article{fu2018morphable,
  title = {Morphable 3D mesostructures and microelectronic devices by multistable buckling mechanics},
  author = {Fu, Haoran and Nan, Kewang and Bai, Wubin and Huang, Wen and Bai, Ke and Lu, Luyao and Zhou, Chaoqun and others},
  journal = {Nat. Mater.},
  volume = {17},
  number = {3},
  pages = {268--276},
  year = {2018}
}

@article{yang2018multistable,
  title = {Multistable kirigami for tunable architected materials},
  author = {Yang, Yi and Dias, Marcelo A and Holmes, Douglas P},
  journal = {Phys. Rev. Mater.},
  volume = {2},
  number = {11},
  pages = {110601},
  year = {2018}
}

@article{haghpanah2016multistable,
  title = {Multistable shape-reconfigurable architected materials},
  author = {Haghpanah, Babak and Salari-Sharif, Ladan and Pourrajab, Peyman and Hopkins, Jonathan and Valdevit, Lorenzo},
  journal = {Adv. Mater.},
  volume = {28},
  number = {36},
  pages = {7915--7920},
  year = {2016}
}

@article{yang2019multi,
  title = {Multi-stable mechanical metamaterials by elastic buckling instability},
  author = {Yang, Hang and Ma, Li},
  journal = {J. Mater. Sci.},
  volume = {54},
  number = {4},
  pages = {3509--3526},
  year = {2019}
}

@article{pandey2014dynamics,
  title = {Dynamics of snapping beams and jumping poppers},
  author = {Pandey, Anupam and Moulton, Derek E and Vella, Dominic and Holmes, Douglas P},
  journal = {Europhys. Lett.},
  volume = {105},
  number = {2},
  pages = {24001},
  year = {2014}
}

@article{medina2014experimental,
  title = {Experimental investigation of the snap-through buckling of electrostatically actuated initially curved pre-stressed micro beams},
  author = {Medina, Lior and Gilat, Rivka and Ilic, Bojan and Krylov, Slava},
  journal = {Sens. Actuators A Phys.},
  volume = {220},
  pages = {323--332},
  year = {2014}
}

@article{feynman1965feynman,
  title = {The {Feynman} lectures on physics; vol. I},
  author = {Feynman, Richard P and Leighton, Robert B and Sands, Matthew and Hafner, Everett M},
  journal = {Am. J. Phys.},
  volume = {33},
  number = {9},
  pages = {750--752},
  year = {1965}
}

@book{landau1976Mechanics,
  title = {Mechanics},
  author = {Landau, Lev Davidovich and Lifshitz, Evgeni{\u{i}} Mikha{\u{i}}lovich},
  volume = {1},
  year = {1976},
  publisher = {Butterworth-Heinemann},
  edition = {Third}
}

@article{gongora2021using,
  title = {Using simulation to accelerate autonomous experimentation: A case study using mechanics},
  author = {Gongora, Aldair E and Snapp, Kelsey L and Whiting, Emily and Riley, Patrick and Reyes, Kristofer G and Morgan, Elise F and Brown, Keith A},
  journal = {iScience},
  volume = {24},
  number = {4},
  year = {2021}
}

@article{hardman2025automated,
  title = {Automated Benchmarking of Variable-Property Soft Robotic Fingertips to Enable Task-Optimized Sensor Selection},
  author = {Hardman, David and Dai, Benhui and Guan, Qinghua and Georgopoulou, Antonia and Iida, Fumiya and Hughes, Josie},
  journal = {Adv. Sci.},
  pages = {e09991},
  year = {2025}
}

@article{tancogne2024using,
  title = {Using miniature experiments to reveal strength gradients in battery casings},
  author = {Tancogne-Dejean, Thomas and Roth, Christian C and Grolleau, Vincent and Beerli, Thomas and Mohr, Dirk},
  journal = {Int. J. Mech. Sci.},
  volume = {275},
  pages = {109253},
  year = {2024}
}

@article{roth2025quantifying,
  title = {Quantifying damage mechanisms through {FE}-based void tracking: Application to shear and tension in-situ laminography experiments on {AA2198-T851}},
  author = {Roth, Christian C and Morgeneyer, Thilo F and Helfen, Lukas and Mohr, Dirk and Tancogne-Dejean, Thomas},
  journal = {Acta Mater.},
  volume = {288},
  pages = {120783},
  year = {2025}
}

@article{meier2024obtaining,
  title = {Obtaining auxetic and isotropic metamaterials in counterintuitive design spaces: an automated optimization approach and experimental characterization},
  author = {Meier, Timon and Li, Runxuan and Mavrikos, Stefanos and Blankenship, Brian and Vangelatos, Zacharias and Yildizdag, M Erden and Grigoropoulos, Costas P},
  journal = {npj Comput. Mater.},
  volume = {10},
  number = {1},
  pages = {3},
  year = {2024}
}

@article{chen2020design,
  title = {Design optimization of soft robots: A review of the state of the art},
  author = {Chen, Feifei and Wang, Michael Yu},
  journal = {IEEE Robot. Autom. Mag.},
  volume = {27},
  number = {4},
  pages = {27--43},
  year = {2020}
}

@article{tong2020microtubule,
  title = {Microtubule simulations provide insight into the molecular mechanism underlying dynamic instability},
  author = {Tong, Dudu and Voth, Gregory A},
  journal = {Biophys. J.},
  volume = {118},
  number = {12},
  pages = {2938--2951},
  year = {2020}
}

@misc{db_placeholder,
  title = {{Experimental Database of Dynamic Instabilities in Bistable Beams}},
  author = {Gutierrez-Prieto, E. and Reis, P.M.},
  publisher = {Zenodo},
  year = {2026},
  note = {Data will be made public and available at this DOI before publication of the article.},
  doi = {10.5281/zenodo.0000000 (Placeholder)},
  url = {https://zenodo.org/record/0000000 (Placeholder)}
}

@inproceedings{paszke2017automatic,
  title = {Automatic differentiation in PyTorch},
  author = {Paszke, Adam and Gross, Sam and Chintala, Soumith and Chanan, Gregory and Yang, Edward and DeVito, Zachary and Lin, Zeming and Desmaison, Alban and Antiga, Luca and Lerer, Adam},
  booktitle = {NIPS-W},
  year = {2017}
}

@article{abbasi2024leveraging,
  title={Leveraging the snap buckling of bistable magnetic shells to design a refreshable braille dot},
  author={Abbasi, Arefeh and Chen, Tian and Aymon, Bastien FG and Reis, Pedro M},
  journal={Adv. Mater. Technol.},
  volume={9},
  number={3},
  pages={2301344},
  year={2024},
  publisher={Wiley Online Library}
}

\end{document}